\documentstyle[aps,preprint]{revtex}
\begin{document}

\title{Cosmological and astrophysical tests of quantum gravity}
\author{Vipul Periwal}
\address{Department of Physics,
Princeton University,
Princeton, New Jersey 08544}

\def\DD{\hbox{D}}
\def\dd{\hbox{d}}
\def\tr{\hbox{tr}}\def\Tr{\hbox{Tr}}
\def\ee#1{{\rm e}^{{#1}}}
\def\part{\partial}
\def\bpart{\bar\partial}
\def\del#1#2{{{\delta #1}\over{\delta #2}}}
\def\refe#1{eq.~\ref{#1}\ }
%\baselineskip=14truept
\maketitle
\tightenlines
\begin{abstract}
{ Physics in the vicinity of  
an ultraviolet stable  fixed point of a quantum field theory
is parametrized by 
a renormalization group invariant macroscopic length scale, the 
correlation length $\xi,$ with the 
quantum effective action a function of this length scale.  
Numerical simulations of quantum gravity 
suggest the existence of just such a fixed point. 
Since the quantum effective action is a function only of $\xi,$ the 
cosmological constant must be $k \xi^{-2}$  with $k$ a pure number.
Higher derivative terms are also parametrized by this length scale, 
so in particular the effective Newtonian dynamics of a test particle is
modified at acceleration scales of order $1/\xi.$  
Thus, renormalization group effects in quantum gravity 
provide a natural link between the 
phenomenological acceleration scale associated with galactic
rotation curves and the value of the cosmological constant favoured by
recent supernovae observations.}

\end{abstract}
\bigskip
Quantum gravity is a difficult subject with a host of conceptual and
computational problems that we are far from resolving.  It is 
the aim of this paper to point out that even in the absence of a complete
theory, general arguments based on the properties of the 
renormalization group in quantum field theory 
suggest that quantum gravity may provide 
simple explanations for some astrophysical and cosmological 
puzzles.  

Let us begin by considering an approach to quantum gravity
reviewed by Weinberg\cite{weinberg}.
Many quantum field theories can be formulated in arbitrary
numbers of spacetime dimensions. Examples are Yang--Mills theories,
the nonlinear sigma model (NLSM), and general relativity (GR).  
In such cases there is
typically a spacetime dimension in which the coupling constant in these
theories is dimensionless, called the upper critical dimension (ucd). For 
the NLSM and GR the ucd is 2 while for Yang--Mills theories
it is 4.  Above the ucd, the coupling constant has
dimensions of length to positive powers and the theory is
perturbatively non-renormalizable---as one probes physics at
shorter distance scales, the effective 
coupling in these theories increases on dimensional grounds 
alone. Suppose that 
exactly at the ucd the  theory is asymptotically 
free\cite{asymp}, in other words quantum loop effects tend to make the coupling 
decrease at shorter distances.  If one  formally treats the dimension
of spacetime as an analytic variable following Wilson and 
Fisher\cite{wilson}, one can find a critical value of the coupling such that
these two effects exactly cancel and the theory is scale invariant at a
dimension $D$ larger than the ucd\cite{zj}. 
This behaviour is encoded in a renormalization group equation
${\partial G/\partial\ln\rho} \equiv \beta(G) = \epsilon G - 
	\beta_{0} G^{2} +\dots $
where $\epsilon\equiv D-D_{c},$ $D_{c}$ is the ucd, 
$G$ is a dimensionless coupling, related to the coupling constant in
$D$ dimensions by $G_{D} = \rho^\epsilon G,$ $\rho$ is the cutoff 
and $\beta_{0}>0$ for asymptotically free theories.  If $D-D_{c}$ is 
small, one can find a fixed point {\it i.e.} $G_{c}:\beta( G_{c}) = 0$
with $G_{c}$ small (in perturbation theory, 
$G_{c}\approx\epsilon/\beta_{0}$) so that one can trust the 
qualitative features found in perturbative computations.  

An important consequence of the 
existence of an ultraviolet stable fixed point is the existence of a
macroscopic physical correlation length $\xi,$ independent of $\rho.$ 
$\xi$  characterizes  scaling violations as one
considers physics at longer distances around the ultraviolet stable 
fixed point, marking the transition from physics 
described by the ultraviolet stable fixed point at $G_{c},$ with 
the effective coupling $G(r/\xi)$ 
growing with increasing $r,$ to physics
described by some other fixed point (in fact, such an infrared fixed point need 
not even exist at finite $G$):
$\xi = \rho \exp\left(\int^{G} dG'/\beta(G')\right) \sim 
	\rho|G-G_{c}|^{1/\beta'(G_{c})}\ .$
While its existence is demonstrable in 
perturbation theory, $\xi$ cannot be determined in perturbation theory.
Fixing $\xi$ specifies $G:$
this is the phenomenon of dimensional transmutation. 

The effective action obtained by systematically computing in 
renormalized perturbation theory is a function of $\xi$ alone.  All 
the terms in the effective action are directly computable 
from the bare action with no arbitrariness.   
One cannot compute  this 
effective action analytically in dimensions far above the ucd 
because the Wilson--Fisher\cite{wilson} $\epsilon$ expansion 
is not  a good quantitative guide, but in principle one 
can compute this effective action numerically.  The only 
features that I need are the  
existence of $\xi$ and the fact that the effective action
is independent of any particular
field configuration.  Its equation of motion incorporates quantum corrections 
and one  must use this quantum  equation of motion for all 
physics with { no change} in
the value of $\xi.$  This may seem surprising but in fact it has to 
be the case since the appearance of $\xi$ is an ultraviolet 
effect---in other words, it is the ultraviolet stable fixed point 
that implies the  existence of $\xi$ and this is independent of
the macroscopic field configuration at which one chooses to evaluate 
the effective action.  

In the present work, I adopt a phenomenological 
stance and attempt to understand  qualitative aspects of physics near 
the ultraviolet stable fixed point in quantum gravity\cite{weinberg}. 
Happily, a more quantitative and constructive 
approach may not be too far in the future. Numerical simulations of
lattice gravity have made great progress and support  the
qualititive features suggested by the $\epsilon$ 
expansion\cite{weinberg}.  I focus on the recent lattice gravity
results of  Hamber and Williams\cite{hw1,hw2}\ in 
the Regge calculus approach\cite{regge}.
For $G>G_{c}$ the effective coupling at a physical separation $r$ behaves as
\begin{equation}
	G(r) = G_{c}\left[1+c_{1} \left({r\over\xi}\right)^{1/\nu} + 
	c_{2}\left({r\over\xi}\right)^{2/\nu}+\dots\right]
\end{equation}
where $\nu\equiv -1/\beta'(G_{c}).$
In the case of pure gravity\cite{hw1,hw2}\ $1/\nu \approx 2.8$ and the
beta function has a negative slope at $G_{c},$ so the effective 
coupling increases with distance.  Further, $1/\xi$ is found to be
very small close to the fixed point, on the order of the inverse
anti--de Sitter radius (since the simulations are Euclidean). 
While the precise value of $\nu$ will  change when matter is added, 
the negative slope of the beta function has been linked to the fact 
that gravity is attractive\cite{hw1}, so the sign of $\nu$ should not change.
  
For general relativity, in the present framework, one expects then that 
the effective action will be a function of a correlation length $\xi$
alone.  Thus, for example, the 
cosmological constant term 
in the effective action will  naturally be $k \int \sqrt g 
(\xi^{2}G_{\xi})^{{-1}}$ with $k$ a computable pure number, and $G_{\xi}$ the 
value of Newton's constant determined non-perturbatively by $\xi.$  
($k$ could be zero but numerical work does not suggest 
this\cite{hw1,hw2}.) Since the effective value of
Newton's constant grows with distance, it is natural to expect that
important aspects of the physics will be incorporated in 
non-polynomial functions of derivatives.   This is exactly analogous 
to Yang--Mills theory where  
physics at large momentum transfer is simple, while at small 
momentum transfer  one sees qualitatively different physics 
characterized by the appearance of the scale 
$\Lambda_{{\rm QCD}}.$  In particular, if
one considers ordinary Newtonian dynamics for a test particle, 
we would expect the appearance of  an acceleration 
scale $1/\xi.$  (The modification of Newtonian dynamics is not due to 
the simple increase in the coupling constant, just as confinement in 
QCD is not due to just the logarithmic growth of the effective 
coupling.)
Lastly, note that string theory does not preclude 
the existence of a non--perturbative 
ultraviolet stable fixed point for quantum gravity
relevant for describing physics at scales much below the string scale.

Why should the scale 
dependence  appear as an acceleration scale?  For example, one might 
imagine that a frequency scale might appear for orbital motion.  
Here one must recall that the effective action's scale dependence is
independent of the geometry under consideration---the scale dependence 
must be something universal that can be defined in gravity for 
{\it any}
geometry.   Based on the  principle of equivalence,
I suggest that the only universal quantity that one can expect is
an acceleration scale.  This is clearly a speculation, but 
at least qualitative confirmation within the $\epsilon$ expansion
should be possible\cite{work}.  Another obvious motivation for an
acceleration scale is provided by a comparison to the  difference 
in small and large momentum transfer processes in QCD.

How do these qualitative features match cosmological and astrophysical 
observations?   I consider cosmology first.  
Recent observational work on supernovae\cite{concord}\ appears to
support a source term in the Friedman equation with 
positive energy and negative pressure, similar to a cosmological 
constant.  The
ratio of the energy in the cosmological constant--like term to
the critical energy density 
$\Omega_{\Lambda}$ is $\approx 0.6-0.7$\cite{concord} for a spatially 
flat universe, which is difficult to explain from a particle physics point of 
view\cite{weinbergcc}.  

If one identifies the Hubble length 
$H_{0}^{-1}=\xi,$ the existence of an ultraviolet stable fixed point in
quantum gravity implies $\Lambda$ is {\it naturally} the same order of
magnitude as $H_{0}^{2}.$ 
Why should the {\it present} value of the 
Hubble length be the correlation length $\xi?$  As 
explained above, the quantum effective action is a function of 
$\xi$ for {\it any} geometry.  While one could  invoke 
anthropic considerations, only within a larger theory can 
one ask this question intelligently, 
just as only within a larger theory can one expect to compute 
the ratio of  $\Lambda_{{\rm QCD}}$ to the mass of the electron. 
The constancy of $\xi$ is important since 
variations of Newton's constant are strongly  constrained\cite{constraints}. 

An explanation for the order of magnitude equality of 
$\Omega_{\rm matter}$ and $\Omega_{\Lambda}$ requires a more complete 
theory as well, see for example\cite{stein}.  
It is interesting to consider the effects 
of this scale on cosmology with  spatial curvature, which is also a viable 
option\cite{rp}. Spatial curvature appears as an integration 
constant in the equations of motion---it is possible albeit
unnatural to have the scale of an integration constant different 
from the scale in the equations of motion.

Now consider galaxy scale physics. As is well--known,  
optical surface brightness in the disks of spiral galaxies falls off 
exponentially with radius.  On the other hand, measured circular 
speeds of rotation as a function of distance from the center of the 
galaxy approach 220 $(L/L_{*})^{0.22}$ km/s\cite{jim}\  
at distances on the order of twice the 
disk length scale.  If starlight traces mass, the measured circular 
velocities should fall off as $r^{{-1/2}}.$  While the 
evidence is most obvious for spiral galaxies, similar puzzles exist 
for other galaxies\cite{jim}.   One can either 
explain this by positing the existence of dark matter, such as
a  preponderance at larger radii of stars with low masses or more 
exotic forms of non--luminous matter, or one can suppose that there
is new dynamics that sets in at these large distances.

There is a large body of phenomenological work 
reviewed in \cite{mond} that is referred to as the modified Newtonian 
dynamics (MOND) 
programme, aimed at explaining observed rotation curves in terms of
modifications of gravity at small  accelerations.
The tenets of MOND are that the actual acceleration of 
a particle ${\bf g}$ is related to the Newtonian acceleration 
${\bf g}_{N}$ by $\mu(g/a_{0}) {\bf g} = {\bf g}_{N} ,$
where $\mu$ is a (matrix) function with the important feature that it
involves a fundamental acceleration $a_{0}.$  Explaining  galaxy rotation 
curves requires that $\mu(x) \rightarrow 1$ as $x\rightarrow \infty,$ 
and $\mu(x) \rightarrow x$ as $x\rightarrow 0.$   
The MOND programme does a good job of explaining the
observed rotation curves\cite{mond}, including the Tully--Fisher 
relation\cite{jim}.  Phenomenology gives 
$a_{0}\approx 2\ 10^{-8}{\rm cm}/{\rm s}^{2}$ which is equivalent to
a length scale of order the present Hubble length, 
a fact that has found  no  explanation\cite{mond}.   There  
are no experimental constraints from terrestrial measurements 
on observed deviations from Newtonian 
dynamics at the accelerations of interest here. Milgrom\cite{mil}\ has
argued already that MOND requires strongly non--local 
higher--derivative terms.

As discussed 
above, non--polynomial higher--derivative corrections
will naturally appear in the quantum effective action with a scale set 
by $\xi.$ While it is not possible to compute these in a quantitative analytic 
form at present, the relation between the 
cosmological constant and the  MOND acceleration scale are 
exactly what one expects in the quantum effective action from the
existence of an ultraviolet stable fixed point.  The small value of 
the acceleration scale is directly related to the large value of the 
correlation length. 
Note particularly in light of the comments above regarding the constancy 
of $\xi$ that 
the light  observed from distant galaxies was emitted at a different
value of the Hubble parameter, but the rotation curves are fit by
an acceleration scale that is equivalent to the {\it present} Hubble 
parameter.  Further, the increase in the
effective Newtonian attraction at larger separations predicted
is qualitatively consistent with the form suggested by phenomenology\cite{mond}.

In conclusion,  I have argued here  
that the dependence of the effective action of  quantum gravity on the 
renormalization group invariant 
correlation length may account for the effective value of
the  cosmological constant and  for the  rotation curves of
galaxies.  There is no way to separate the cosmological implications and
the implications for small accelerations of the appearance of 
$\xi$---quantum field theory  predicts a universal scale parameter.  
The relations suggested in the present letter are conservative and 
can be tested numerically.  The qualitative 
features of quantum gravity are in analogy with the proven physics of the 
NLSM, and are supported by the available lattice data.  In ending     
I must mention that quantum gravity has been suggested previously 
as a source of infrared 
effects---see for example \cite{woodard}.  The connections 
considered here have not been suggested in the literature---see, 
however, \cite{mannheim}\ for other work connecting a universal acceleration 
scale and a cosmological constant.

I am grateful to J. Cohn, H. Hamber, 
G. Lifschytz, M. Milgrom, J. Peebles, L. Randall, B. Ratra, P. 
Steinhardt, C. Thompson and M. White for their help.
This work was supported in part by NSF grant PHY-9802484.

%Correspondence to Vipul Periwal, vipul@mail.princeton.edu.
\end{document}